# Data reduction procedure for correction of geometrical factors in the analysis of specular x-ray reflectivity of small samples

**Arijeet Das[a]\*, Shreyashkar Dev Singh[a], R. J. Choudhari[b], S. K. Rai[a] and Tapas Ganguli[a]**

[a] Raja Ramanna Centre for Advanced Technology, RRCAT, Indore, Madhya Pradesh, 452013, India
[b] UGC-DAE Consortium for Scientific Research, Indore, Madhya Pradesh, 452017, India

Correspondence email: arijeet@rrcat.gov.in

**Synopsis**   A data reduction procedure for the removal of the effect of geometrical factor from the angle dependent specular x-ray reflectivity profile of small samples is described.

**Abstract**   For small samples, the modification of the XRR profile by the geometrical factors manifesting due to profile and size of the beam and the size of the sample is significant. Geometrical factors extend till spill over angle which is often greater than critical angle for small samples. To separate the geometrical factor, it is necessary to know the spill over angle. Since geometrical factor is a smoothly varying function and extends beyond critical angle, it is impossible to determine the spill over angle from XRR profile of small samples. We have shown by comparing the normal XRR profile of a small sample with the XRR profile taken with a surface contact knife edge on the same sample, that the spill over angle can be determined. Thus we have developed a procedure for data reduction for small samples and validated it with suitable experiments. Unlike hitherto used methods which have drawbacks, this is a self consistent method for data reduction.

## 1. Introduction

Specular x-ray reflectivity (XRR) is a non destructive technique for characterization of flat surfaces or thin films deposited on a flat surface. It mostly determines electron density profile normal to the surface which is related to the thickness of different layers and roughness at their respective interfaces. Determining such information without physically damaging or altering the sample makes the technique attractive for thin film characterization. The theory of XRR is primarily contained in Fresnel's law for reflection of electromagnetic waves at ideal sharp and flat interfaces defined by change of refractive index. Layer(s) of thin film(s) on a substrate offer multiple interfaces. The reflected x-ray beam from different interfaces, owing to their difference in phase which depends on refractive index of different layers and path length, produce Bragg like reflection peaks in the XRR profile due to interference effect. Physical roughness, chemical roughness or intermixing between layers reduces the contrast at the interfaces due to a relatively gradual variation (transition) of density in real interfaces.

Parratt's recursive formalism is the procedure to calculate the XRR profile from thin film samples by application of Fresnel's law for sharp interfaces **(Parratt, 1954)**. Novet and Croce model incorporates the effect of roughness by multiplying a factor of $exp(-q_z^2 \sigma^2 / 2)$ to the x-ray reflectivity where qz is the vertical wave vector transfer and σ is the roughness in angstroms **(Nevot and Croce, 1980)**. Combination of these two is central and sufficient to describe XRR completely.

Conventionally, XRR is measured experimentally by a θ-2θ scan which records the reflected intensity of monochromatic x-ray beam from a thin film sample as a function of 2θ. The experimentally obtained curve is not exactly the same as the Fresnel reflectivity for the following two reasons. The first one is due to diffuse scattering which scatters some x-ray photons in the specular direction. However the diffuse scattering background is not significant for samples having uncorrelated roughness and the method to separate it in the relevant situations is described in literature **(Romanov et.al., 2010)**. The second reason is due to the experimental configuration, coined as the geometrical factors. It includes modifications in the specular XRR profile due to size of the sample and size and profile of the incident x-ray beam. The usual practice is to suitably modify the experimentally observed XRR profile by various data reduction procedures described in the literature and then to fit it to the theoretical curve calculated using Parratt's recursive formalism to extract numerical values of thickness, roughness and density to different layers. All the available XRR analysis softwares generate Fresnel's reflectivity (modified by roughness factor) to compare to the experimentally observed XRR profile. The primary function of the data reduction procedures is to remove the effects of geometrical factors and extract the precise Fresnel's reflectivity from the experimentally observed curve. Discussion on whether such a procedure can actually generate the precise Fresnel's reflectivity is lacking in the available literature. The criteria suggested by Gibaud et al. to reduce complications in data reduction procedure is to use larger samples and judiciously select the size of entrance slit such that :

$$L > (T / \sin \theta_c) \quad \text{............................(1)}$$

where L is the length of the sample, T is the aperture of the front slit and $\theta_c$ is the critical angle of the sample **(Gbaud et.al. 1993)**. When equation (1) does not hold, the XRR profile gets significantly modified by geometrical factors which is also known as foot print effect arising due to the size and profile of the beam and size of the sample. Reducing the width of entrance slit (T) has limitations. **(Salah et. al. 2017)**. At very small apertures, the size of the beam at the sample position becomes poorly defined in terms of size and divergence in addition to loss of flux. Furthermore all thin film deposition methods cannot produce large samples with desired quality. For example in pulsed laser

deposition(PLD), a widely used technique for thin film deposition due to its versatile material selectivity and tuneability of deposition conditions, the maximum size with which samples can be deposited with desired thickness uniformity is ~ 10mm x10mm. This makes PLD grown samples unsuitable for XRR characterization in the absence of proper data reduction procedure and more so when the thin films have smaller value of $\theta_c$. Hence in order to extend the scope of applicability of XRR, it is of vital interest to develop a suitable method for correction of geometrical factors for situations where it is difficult to satisfy equation (1). It has been proposed in the literature to calculate the geometrical factor by calculating the derivative of the reflectivity curve before the critical angle. In this work we have argued that this method can be misleading. Traditionally the XRR curve is normalized by using the direct beam count or by using an initial guess of the average density of the sample. We have discussed in the context of reference **(Salah et. al. 2017)** and **(Gbaud et.al. 1993)** that both these methods are not correct.

Even when equation (1) holds, as we have discussed in later sections, the experimental XRR profile is not precisely same as Fresnel's reflectivity. In this paper we have addressed these issues by developing procedures to separate geometrical factors from experimental XRR. These procedures in general are applicable for samples of all sizes but especially useful for small samples where data reduction is critical. Our objective is to develop a data reduction procedure which is applicable to all sample sizes and it should not use any of the direct beam count, derivative of the XRR curve, physically measured values of sample size or prior knowledge of the optical constants of the thin film material to normalize the curve.

## 2. Theory

In this section we elaborate certain important aspects of the experimental geometry of XRR which will be used for data reduction. The angular dependence of calculated Fresnel reflectivity will exactly correspond to the experimentally obtained reflectivity curve, if the flux available for reflection is constant at all angles. The x-ray source provides constant flux with finite and constant beam width and a fixed profile. However in the experimental geometry of the θ-2θ scan, the angle made by the sample of finite size with the beam propagation direction starts from θ=0° with the height of the sample so adjusted that it cuts half of the flux from the source. At θ=0° the sample being parallel to the beam, does not reflect at all. As θ increases, the sample surface starts intercepting the beam depending on its finite size. This is the reason for a monotonically increasing flux available for reflection with increase of angle. The extent of interception of the x-ray beam by the sample is given by ($L\sin\theta$) where L is the length of the sample and θ is the glancing angle of incidence. The intercepted beam Φ(θ) is available for reflection. This can be express as:

$$\Phi(\theta) = \int_{(-L/2)\sin\theta}^{(L/2)\sin\theta} P(z)dz \quad \ldots\ldots\ldots\ldots\ldots\ldots(2)$$

where P(z) is the beam profile. If T is the vertical size of the beam then the full beam falls on the sample when θ is such that $L\sin\theta = T$. This angle is called spill over angle $\theta_{so}$. For all angles $\theta > \theta_{so}$ there is no variation in the flux available for reflection and the experimental reflectivity, if normalized correctly, is precisely equal to calculated Fresnel's reflectivity. For all angles $\theta \leq \theta_{so}$ the experimental curve is the Fresnel's reflectivity modulated by $\Phi(\theta)$ given by equation (2). This modulation is known as geometrical factor or foot print effect.

**2.1. Geometrical factor due to beam and sample size:**

Fresnel's reflectivity R(θ) of a flat surface is given by:

$$R(\theta) = \frac{\left|\theta - \sqrt{\theta^2 - 2\delta + 2i\beta}\right|^2}{\left|\theta + \sqrt{\theta^2 - 2\delta + 2i\beta}\right|^2} \quad\ldots\ldots\ldots\ldots\ldots(3)$$

where δ and β are the real and imaginary part of refractive index of the sample respectively.

For simplicity, let us consider that the beam profile P(z) is a rectangular step function of width T and height P(z) = 1. With $L\sin\theta_{so} = T$ and from equation (1) we get:

$$\Phi(\theta) = L\sin\theta, \quad for\ \theta < \theta_{so}$$
$$= T, \quad for\ \theta \geq \theta_{so} \quad\ldots\ldots\ldots\ldots(4)$$

Equation (4) can be normalized by dividing beam width T to get the geometrical factor $g(\theta)$.

$$g(\theta) = (L/T)\sin\theta, \quad for\ \theta < \sin^{-1}(T/L)$$
$$= 1, \quad for\ \theta \geq \sin^{-1}(T/L) \quad\ldots\ldots(5)$$

Where, $\sin^{-1}(T/L) = \theta_{so} \ldots\ldots\ldots\ldots\ldots\ldots(6)$

In figure 1 we have plotted the geometrical factor g(θ) for different values of (L/T). It can be seen in the figure that the spill over angle increases with decreasing (L/T) ratio. Experimental XRR profile *E(θ)* can be calculated by multiplying equation (4) to equation (5). In figure 2 we have plotted *R(θ), g(θ) and E(θ)* [= *R(θ)g(θ)*]. *R(θ)* is calculated using δ=2.145x10$^{-05}$ and β= 1.262x10$^{-06}$. It can be observed that *R(θ)=1* for θ=0º. In the region of total external reflection (from $\theta = 0º$ to $\theta = \theta_c^o$), *R(θ)* decreases slowly and monotonically due to absorption factor *β* of the sample **(Parratt, 1954)**. This feature in not observed in *E(θ)* curve which is the result of modification of *R(θ)* by *g(θ)*. The XRR measurement produces *E(θ)*, and the goal of a data reduction procedure is to produce *R(θ)* from *E(θ)*.

It can be seen in figure 2 that for $\theta > \theta_{so}$, the Fresnel's reflectivity $R(\theta)$ and experimental curve $E(\theta)$ merge with each other.

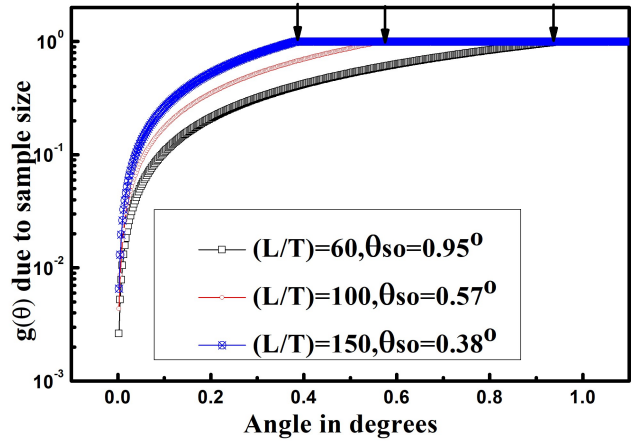

**Figure 1** Plot of $g(\theta)$ in log scale calculated for (L/T)=60, 100 and 150. The values of $\theta_{so}$ in degrees are mentioned in the figure.

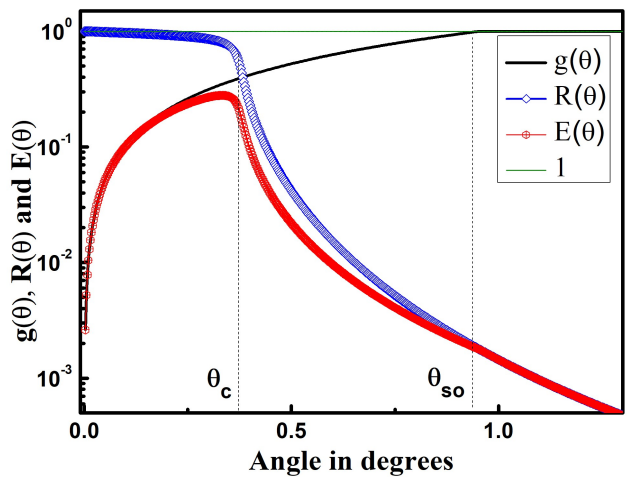

**Figure 2** Calculated $R(\theta)$, $g(\theta)$ and $E(\theta)$.

The actual reflectivity curve contains a part of the direct beam at the values of $\theta$ close to $0^o$. Before starting the $\theta$-$2\theta$ scan, the usual practice is to align the sample in such a way that it obstructs the lower half of the direct beam. Hence during the onset of the scan, the measured intensity as $\theta$ starts from $0^o$ is the sum of two components. The first one is the unobstructed half beam intensity which is being progressively reduced by the increasing obstruction offered by the sample due to increase in $\theta$. The second component comes from the reflection of a portion of the initially obstructed lower half of the beam by the sample as $\theta$ increases. Furthermore it is well-known that the profile of the direct beam obtained from such a scan shows a shift in the angular position of the source due to chopping off of the lower half portion of the beam. It is also reported that recording the half of the incident flux with the sample parallel to the beam in the half cut positions is also not free form errors. Usually the alignment procedures are so critical that the error is often ~10% or larger **(Gbaud et.al.**

**1993)**. Hence we argue that the intensity of the direct beam obtained from the θ-2θ scan cannot be precisely modelled and parameterized in relation to the actual parameters of the source. For this reason we have not used the maximum intensity of direct beam obtained from the experimental curve for any of our modelling and subsequent calculations on actual experimental data. Thus we have demonstrated that the information contained in the beam which is actually reflected form the sample (direct beam is not reflected from the sample) is sufficient to normalize the reflectivity curves. This eradicates the effect of alignment errors.

**2.2. Effect of the beam profile on the geometrical factor.**

In this section we have considered that the beam profile of the x-ray source can be a rectangular step function, or a Gaussian or a super Gaussian described by equations 7 to 9 respectively and shown in the figure 3. The profile functions are:

$$P(z) = 1, \quad for \quad |z| < \sigma$$
$$= 0, \quad for \quad |z| > \sigma \quad \ldots\ldots\ldots\ldots(7)$$

$$P(z) = \frac{1}{T} \exp\left(-(z^2/2\sigma^2)\right) \ldots\ldots\ldots\ldots\ldots\ldots(8)$$

$$P(z) = \frac{1}{T} \exp\left(-(z/\sqrt{2}\sigma)^n\right) \ldots\ldots\ldots\ldots\ldots\ldots(9)$$
$$where \quad n = 4, 6, 8\ldots$$

σ is the half width at half maximum (HWHM) of the beam and T is the width of the entrance slit. In practice, σ = T/2 **(Gbaud et.al. 1993)**.

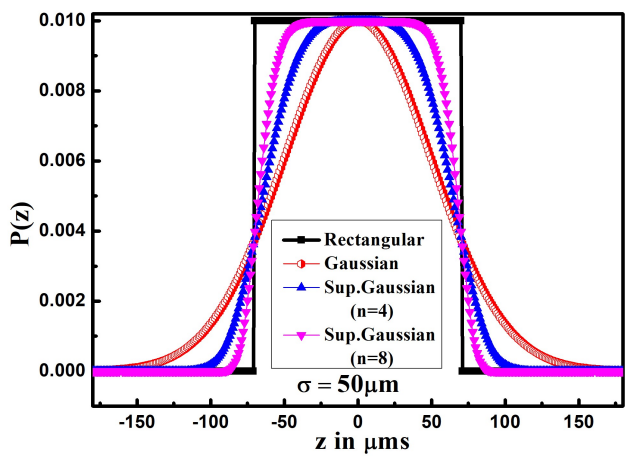

**Figure 3** Rectangular, Gaussian and supper Gaussian beam profiles with order n=4 and n=8.

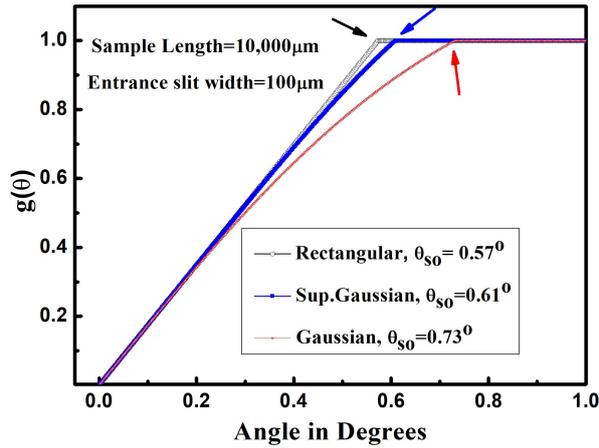

**Figure 4** Calculated foot print function g(θ) for rectangular, super Gaussian of order 4 and Gaussian beam profiles. The sample length and entrance slit width used for calculations are 10,000μm and 100μm respectively. The values of spill over angle obtained from the calculations are $0.57^0$, $0.61^0$ and $0.73^0$ respectively.

The foot print function g(θ) can be calculated from P(z) as follows:

$$g(\theta) = \int_{-(L/2)\sin\theta}^{(L/2)\sin\theta} P(z)dz \bigg/ \int_{-(L/2)\sin\theta_{so}}^{(L/2)\sin\theta_{so}} P(z)dz \quad\ldots\ldots\ldots(10)$$

It may be noted that equation 10 reduces to equation 5 for rectangular beam profile of width T. Figure 4 shows the plot of foot print function g(θ) for different beam profiles. The length of the sample and the entrance slit width used in the calculation are 10,000 μm and 100 μm respectively. It is observed that the slopes of the foot print functions changes according to the type of beam profile. Hence the method described by **(Baules et.al. 2006)** to calculate the sample length from the slope of *E(θ)* below the critical angle is only an approximation and more so when we appreciate that this portion of the experimental curve is also multiplied with *R(θ)* which is a monotonically decreasing function due to finite absorption factor *β* **(Baules et.al. 2006)**. For lower angles, the slopes of the function *g(θ)* remain same for all the three type of beam profiles. But at higher angle, they differ. Hence if the critical angle $\theta_c$ of the sample is small, its evaluation by using the foot print function for different beam profiles will not vary significantly. For the Gaussian beam profile the curvature of the plot is prominent and visible. It can be observed from the figure that the Gaussian beam profile shows highest spill over angle ($\theta_{so}$). As the beam profile approaches the rectangular box function, $\theta_{so}$ gradually decreases.

The awareness that the spill over angle $\theta_{so}$ is not only dependent on the physical size of the sample and the beam but also on the beam profile, leads to the conjecture that it cannot be calculated by simply measuring the physical parameters L and T. As can be seen in figure 4, the shift in spill over angle for different beam profiles is as much as $0.16°$. Hence it would be incorrect to calculate it from physically measured values of L and T by using equation 6 as suggested in reference **(Salah et. al. 2017)**.

**2.3. Classification on the basis of the span of geometrical factor.**

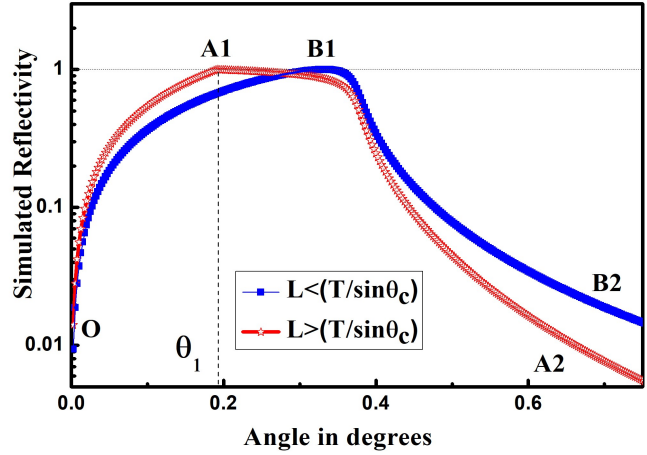

**Figure 5** Calculated reflectivity of the same sample with different sample length

Figure 5 shows calculated reflectivity for two cases. For the curve OA1A2, $\theta_{so} < \theta_c$ and for curve OB1B2, $\theta_{so} > \theta_c$. For reasons described in section 2.1, the curves are normalized to their respective maximum instead of the direct beam maximum. We observe that in the curve OA1A2, the geometrical factor is present from O to A1 and the rest of the curve from A1 to A2 is parallel to Fresnel's reflectivity. Since its value at A1 is 1 which ideally should have been less than 1 to account for finite absorption $\beta$ in the sample; it can be mathematically represented as $A1A2 = R(\theta)/R(\theta_1)$. Available XRR fitting programs calculate $R(\theta)$ and compare it to the experimentally observed XRR. However due care has to be taken when the absorption factor $\beta$ is large and $\theta_{so}$ is very close to but less than $\theta_c$. In practice, prior knowledge of $\delta$ and $\beta$ of the sample are used to generate $R(\theta)$ and then the experimental curve is multiplied with a suitable factor to match with $R(\theta)$ in such situations. Deposition conditions alter density of the samples. There is a fair chance that this change may go undetected due to subjective biasing introduced into the fitting procedure by using prior knowledge of $\delta$ and $\beta$ to decide the multiplication factor. Hence in such cases, for the sake of theoretical precision, we fit the experimental curve to $R(\theta)/R(\theta_1)$ instead of fitting it to $R(\theta)$. Here $\theta_1$ is the angle at which experimental reflectivity profile is maximum.

In the curve OB1B2 the geometrical factor is present beyond B1. It is impossible to detect the spill over angle. The data reduction procedure developed in this work is categorically useful for this situation. However it should be noted that the procedure developed here will be applicable if $\theta_{so}$ is even greater than the maximum range of angle $\theta$s up to which experimental reflectivity signal is measurable.

### 2.4. θ-2θ scan with a surface contact knife edge

In several commercial reflectometers, beam knife edges are routinely used for reducing the beam foot print and reduce air scattering. For small samples it can be used primarily to reduce the beam foot print. However due to difficulties involved in the process of alignment it is not easy to check the parallelism of the knife edge with the sample and it is also difficult to know the exact distance between the knife edge and sample. We use a knife edge made up of normal paper cutting blade and make it stand on the sample with its own weight by a simple mechanical arrangement. The weight of our knife edge is within 5 grams and is far less heavy than usual commercial knife edges.

The average distance between the knife edge and the sample is nominal $t_{ke} \approx 2 \mu m$. Since the nominal sample size is ~10,000μm, the (L/T) ratio obtained with this setup is sufficient to yield spill over angle far less than critical angle.

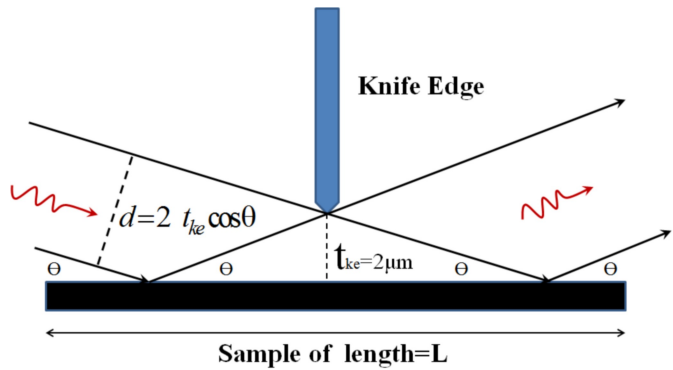

**Figure 6** Schematic setup for measurement with a surface-contact beam knife edge

In the figure 6 we have drawn the schematic diagram of the surface contact knife edge arrangement. The allowed beam width can be calculated to be $2 t_{ke} \cos\theta$. However, taking θ-2θ scan with this configuration shows that the effect of finite absorption before critical angle is not observed in the XRR profile (please refer to $E_{KE}(\theta)$ of figure 8). This may be occurring due to small changes in the beam size and divergence which can be dependent on the slope error of the sample or small misalignments which are unavoidable at these small angles. Due to this distortion occurring before critical angle this XRR profile cannot be normalized precisely. But due to high (L/T) ratio, if normalized correctly, it will exactly coincide with theoretical reflectivity after critical angle none the less.

### 2.5. Procedure of data reduction

In this sub-section we have described a generalized form of data reduction procedure for arbitrary beam profile. In section 4 we have elaborated for the specific case when the beam profile is a step function.

Our data reduction procedure uses two experimental reflectivity profiles of the same sample. The first XRR profile is taken without any knife edge and is denoted by E(θ). The second XRR profile is obtained with a surface contact knife edge denoted by $E_{KE}(θ)$. The Fresnel's reflectivity R(θ) is the unknown quantity to be evaluated from these information. We assume that E (θ) and $E_{KE}(θ)$ can be expressed as:

$$E(\theta) = g(\theta)R(\theta)C_1 \quad \text{...............(11)}$$

$$E_{KE}(\theta) = g_{KE}(\theta)R(\theta)C_2 \quad \text{...............(12)}$$

Where $g(\theta)$ and $g_{KE}(\theta)$ are the geometrical factors given by equation (5), $C_1$ and $C_2$ are the constants arising due to normalization. Since spill over angle in $E_{KE}(θ)$ is very small we can consider $g_{KE}(\theta) = 1$ for all practical purposes. Hence, from equation (12) we get:

$$E_{KE}(\theta) = R(\theta)C_2 \quad \text{...............(13)}$$

We define $E'(\theta)$ such that $E'(\theta) = E(\theta)/(C_1/C_2)$; thus from equation (11):

$$E'(\theta) = g(\theta)R(\theta)C_2 \quad \text{...............(14)}$$

Writing $g(\theta)$ explicitly:

$$g(\theta) = \frac{1}{T} \int_{(-L/2)\sin\theta}^{(L/2)\sin\theta} P(z)dz, \quad \text{for } \theta < \sin\theta_{so}$$
$$= 1, \quad \text{for } \theta \geq \sin\theta_{so} \quad \text{........(15)}$$

Where P(z) can be defined by any one of the equations (7),(8) or (9). In order to calculate g(θ) it is necessary to find the spill over angle. We compare E'(θ) and $E_{KE}(θ)$ to find out the spill over angle. From equation (15) for $\theta \geq \theta so$, $g(\theta) = 1$; thus by equation (13) and (14);

$$E'(\theta) = E_{KE}(\theta) \text{ for } \theta \geq \theta so \quad \text{...............(16)}$$

Equation (16) suggests that $E'(\theta)$ and $E_{KE}(\theta)$ meet each other at $\theta = \theta so$. Thus θso is determined. Satisfying condition that;

$$\frac{1}{T} \int_{(-L/2)\sin\theta_{so}}^{(L/2)\sin\theta_{so}} P(z)dz = 0.996 \quad \text{...............(17)}$$

the value of the effective sample length L can be calculated. The above condition guarantees that the sample covers 99.6% of the beam at θso which corresponds to $\pm 3\sigma$ of the beam profile. (For rectangular step function like beam profile, the above condition reduces to $(L/T)\sin\theta so = 1$, from where the effective value of (L/T) can be calculated). Now g(θ) can be calculated for all values of θ by using the values of L and θso in equation (15). We identify $\theta_1$ as the starting value of glancing angle and write $R(\theta_1)$ as $R_1$ which is very close to 1. Then we calculate from equation (11) & (15);

$$R_1 C_1 = E(\theta_1)/g(\theta_1) \quad\quad\quad\quad\quad (18)$$

$$R_1 C_2 = R_1 C_1 / (C_1 / C_2) \quad\quad\quad\quad\quad (19)$$

Using equation (11) & (18) we calculate;

$$\Psi(\theta) = R(\theta)/R_1 = E(\theta)/[g(\theta) R_1 C_1] \quad\quad (20)$$

And using equation (12) & (19) we calculate;

$$\Psi_{KE}(\theta) = R(\theta)/R1 = E_{KE}(\theta)/[R1 C_2] \quad\quad (21)$$

Since $R_1 = R(\theta_1) \sim 1$, both $\Psi(\theta)$ and $\Psi_{KE}(\theta)$ can be used independently used for fitting with Fresnel's reflectivity. For the sake of generality and theoretical accuracy, we fit $\Psi(\theta)$ and $\Psi_{KE}(\theta)$ with $R(\theta)/R_1$.

It should be noted that if L and T are the physically measured dimensions of the sample and the beam respectively, then the maximum value of θso considering a Gaussian beam profile can be calculated by the following formula for a quick approximation.

$$\theta_{so}^{max} = \sin^{-1}(1.26 T / L) \quad\quad\quad\quad\quad (23)$$

### 3. Experimental

#### 3.1. Sample Preparation

We have deposited $Er_2O_3$ thin film samples on GaAs substrate by pulsed laser deposition (PLD) technique using Excimer laser of wavelength 248nm. The dimension of the substrates was kept at 10mmX10mm. It's a well know fact that desirable thickness uniformity is not achieved by PLD if bigger substrates are used for deposition. Five samples were deposited by varying the substrate temperature and ambient oxygen pressure during deposition. Laser energy, repetition rate and deposition time were similar for all the samples. The details of the samples are listed in table 1.

**Table 1** Deposition environment of various samples.

| Sample Name | Substrate temperature in oC | Chamber pressure in mbar |
|---|---|---|
| S 1 | 550 | 9x10-5 |
| S 2 | 600 | 9x10-5 |
| S 3 | 650 | 9x10-5 |
| S 4 | 550 | 5x10-3 |
| S 5 | 550 | 5x10-2 |

## 3.2. Measurements

Normal XRD measurements in θ-2θ geometry were done for all the samples. It was found that the samples were crystalline and variation of deposition temperature has no visible effect on the crystal structure of the material till ambient oxygen pressure was maintained below $5 \times 10^{-3}$ mbar. However when oxygen partial pressure was increased to $5 \times 10^{-2}$ mbar, the (222) peak of $Er_2O_3$ shifted to lower 2θ value which corresponds to the actual lattice parameter of bulk Er2O3. This suggests that the lattice contracted for lower oxygen partial pressure.

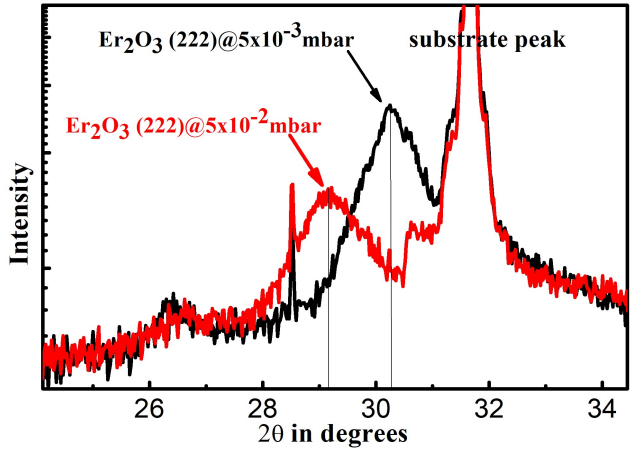

**Figure 7** XRD pattern of the sample deposited at $5 \times 10^{-3}$ mbar pressure is plotted in black and that of the sample deposited at $5 \times 10^{-2}$ mbar is plotted in red.

Figuer 7 shows the change of lattice parameter of Er2O3 layer in sample 4 and 5, due to change in ambient oxygen pressure. This change in lattice parameter must reflect in change in density of the samples which should be detectable in the XRR measurements. With the method of data reduction described in section 2, we carried out the XRR analysis of all the samples. The hard x-ray characterizations of all the samples were done using the grazing incidence x-ray reflectivity (XRR) technique on a Bruker Discover D8 reflectometer and diffractometer with Cu Kα radiation (λ = 1.54 Å) and using a $K_\beta$ filter. The θ and 2θ values were aligned to better than $0.002^0$. Specular reflectivity was measured in θ–2θ geometry over the range of 0° to 4° with a step size of $0.005^0$ for all the samples. As described in section 2.4, we also recorded the XRR measurements on the samples using surface contact knife edge.

A program was developed for data reduction in MATLAB. The processed data were then used for reflectivity fitting using Parratt32 reflectivity fitting software.

## 4. Analysis

The direct beam part of the normal XRR profile and the surface contact knife edge XRR profiles were removed and then they were normalized with respect to their respective maximum. They are plotted in figure 8 and demarcated by $E(\theta)$ and $E_{KE}(\theta)$ respectively. We assume a step function like beam profile described in equation 7.

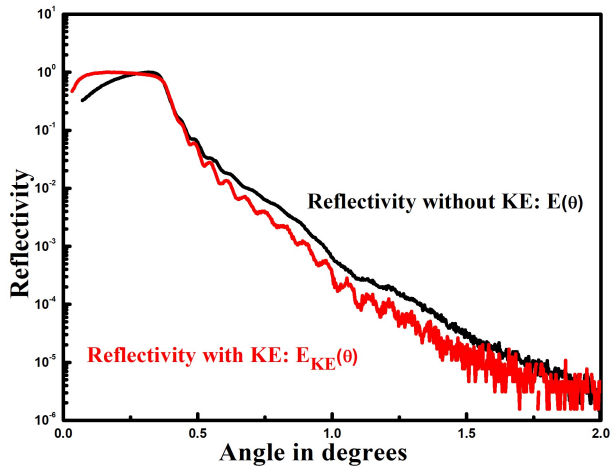

**Figure 8** Reflectivity measured with and without Knife edge plotted against angle.

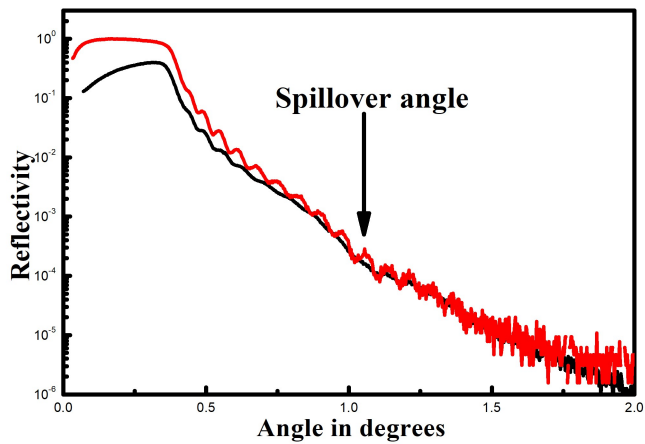

**Figure 9** E(θ) (black curve) is divided by a constant to be identified as (C1/C2) so that the high angle part of both the curves coinside

Referring to figure 8, we identify the x-coordinate of the first data point of $E(\theta)$ as $\theta_1$. The y-coordinate of the first data point of $E(\theta)$, by equation 11 is:

$$E(\theta_1) = g(\theta_1)R_1C_1 \ldots\ldots\ldots\ldots(23)$$

Here we identify the Freshnel reflectivity at $\theta_1$ to be $R_1$.

As discussed in section 2, $E_{KE}(\theta)$ is same as $R(\theta)$ for higher angle. We divide $E(\theta)$ by a suitable factor to obtain $E'(\theta)$ so that $E_{KE}(\theta)$ and $E'(\theta)$ coincide at higher angle. We identify the factor to be (C1/C2). In figure 9 we have plotted both the curves after performing this operation. The angle at which both these curves meet is identified as the spillover angle.

We apply $(L/T)\sin\theta_{so} = 1$; to calculate (L/T). $g(\theta)$ is calculated from equation (5) by using the value of (L/T). Thus the value of $g(\theta_1)$ is found which is used in equation (18) to find out the value of R1C1. Then R1C2 can be found by equation (19).

$\Psi(\theta)$ and $\Psi_{KE}(\theta)$ are determined by using equations 18 to 21 and the values of $g(\theta)$, R1C1 and R2C2. We have plotted $E(\theta)$ and $\Psi(\theta)$ in figure 10. It can be observed that $\Psi(\theta)$ which is derived from $E(\theta)$ shows the monotonically decreasing trend which is the effect of finite absorption before the critical angle.

It is worthwhile to study the behaviour of the function $g(\theta)R_1C_1$ with respect to the spill over angle $\theta_{so}$. Change in $\theta_{so}$ changes $g(\theta)$ through equation 5 and 6. So, following equation 5, we write:

$$g(\theta)R_1C_1 = E(\theta_1)\sin\theta / \sin\theta_1, \quad \text{for } \theta < \theta_{so}$$
$$= R_1C_1, \quad \text{for } \theta \geq \theta_{so} \quad \ldots\ldots\ldots(25)$$

Let us consider the case: $\theta < \theta_{SO}$. In the RHS of the above equation, $E(\theta_1)$ and $\theta_1$ are constants taken from the experimental curve. They are independent of $\theta_{so}$. So the RHS is independent of spiller over angle. But in the LHS, $g(\theta)$ is calculated from $\theta_{so}$. If there is any uncertainty in evaluation of the spill over angle, there will be uncertainty in the value of $g(\theta)$. In such situation, to satisfy LHS=RHS, the factor $R_1C_1$ in the LHS will take an appropriate value. In other words $g(\theta)$ and $R_1C_1$ conspire to keep the product $g(\theta)R_1C_1$ a constant independent of spill over angle. It is this feature of the procedure that makes it appropriate for data reduction even when the spill over angle is greater than the maximum angle up to which data can be measured (i.e. for very small samples sizes). The function $g(\theta)R_1C_1$, for different values of spill over angle are plotted in figure 10. It can be observed that this function is unique for $\theta < \theta_{SO}$ for all values of $\theta_{SO}$.

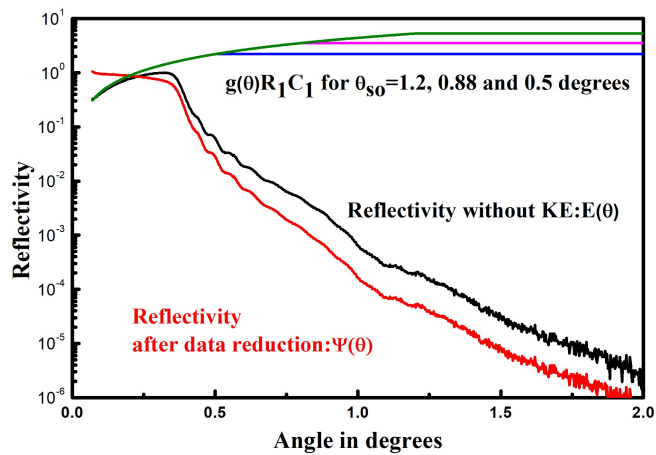

**Figure 10** E(θ) and Ψ(θ) are plotted. The curves of g(θ)R1C1 for different values of spill over angle are also plotted. The spill over angle used for calculation of Ψ(θ) is 0.88 degrees.

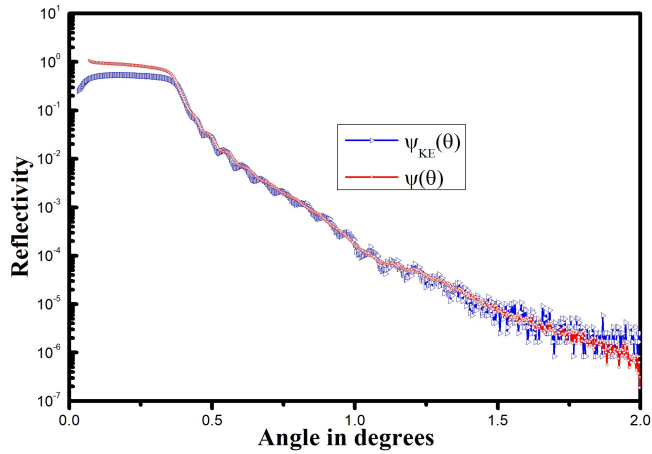

**Figure 11** The normalized reflectivity curves Ψ(θ) and ΨKE(θ) are plotted.

Figure 11 shows the final normalized curves after data reduction. The constant $C_2$ is related to the spill over angle through equations 6, 18 and 19. Both these curves can be fitted with the Fresnel reflectivity to produce identical results. The mismatch of these curves before critical angle shows that the knife edge modifies the XRR profile before critical angle.

**5. Results and discussion**

Reduced data for all the five samples were taken for fitting in Parratt32 reflectivity fitting software. The XRR with the knife edge show clear thickness oscillations where as these are smeared out in the XRR profile without knife edge. The reason for this is the increased resolution of the measurement due to reduction of beam width by the knife edge. We have confirmed by fitting both these data that both of them correspond to same density variation model with different experimental resolution. The fitting are shown in figure 12. In table 2, we have mentioned the fitting parameters applicable for both curves.

**Table 2** XRR fitting parameters

|           | Thickness In angstroms | Real part of ρ | Imaginary part of ρ | Roughness in angstroms |
|-----------|------------------------|----------------|---------------------|------------------------|
| Interface | 93.13                  | 5.5E-6         | 1.6E-7              | 7.6                    |
| $Er_2O_3$ | 643.6                  | 5.8E-5         | 4.7E-6              | 6.7                    |
| Interface | 61.04                  | 3.9E-5         | 5.2E-6              | 25.5                   |
| substrate | Inf                    | 3.8E-5         | 1.1E-6              | 4.1                    |

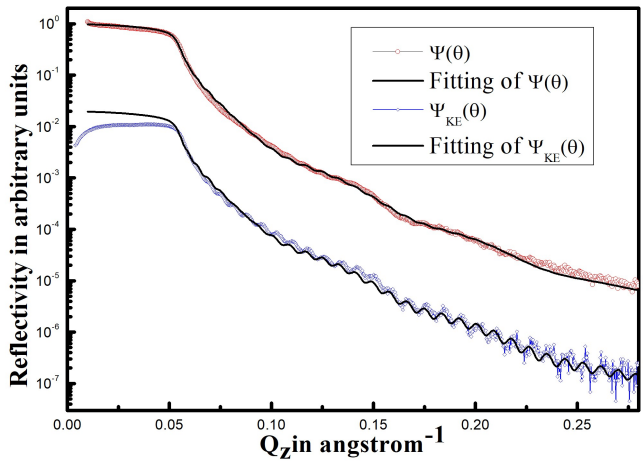

**Figure 12** Fitting of $\Psi(\theta)$ and $\Psi KE(\theta)$. Both of them fit with same density variation model but with different experimental resolution. So, both of them represent same micro structural information.

Sample 5 was deposited at $5 \times 10^{-2}$ mbar pressure and the XRD measurements shown in figure 7 suggest that there was an increase of lattice parameter for this sample. The Φ-scan in XRD measurements ( graph not shown here) confirm that the film was not epitaxial. Hence it has no relation to the lattice parameters of the substrate which indicates there is no biaxial strain in the sample. So this change in lattice parameter must correspond to a change in density of the sample. The scattering length density from XRR measurements shown in figure 13 also show that the density of this sample has decreased by ~6% than the other four samples. Furthermore it is worthwhile to mansion that the density of samples S1 to S4 are ~4.6% higher than the actual bulk density of $Er_2O_3$ where as density of sample 5 is 1.6% lower. The detection of this change of density would not have been possible by the prevailing method of data normalization where the experimenter uses the available value of δ and β to normalize the XRR profiles and thus introduces an unwanted subjective bias into the fitting procedure. The small change of density due to variation of deposition parameters would have gone unnoticed.

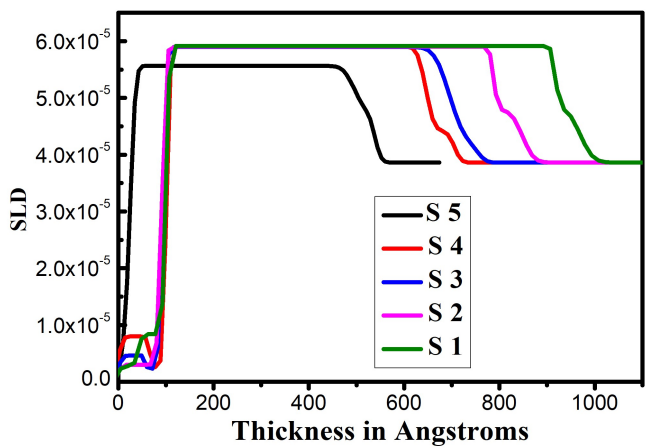

**Figure 13**   Scattering length Density variation modes obtained from fitting of XRR data of the Er2O3samples.

We have also performed the data reduction assuming the beam profiles to be a Gaussian or super Gaussian described by equations 8 and 9. They yield similar result. The difference in the deduced values of δ and β were negligible. This is as a result of insignificant difference between g(θ) values calculated for these beam profiles at lower angle as is evident in figure 4.

## 6. Conclusion

We have proposed a method of data reduction for the XRR measurement of small samples. We have not used direct beam count, derivative of the XRR curve, physically measured values of sample size or prior knowledge of the optical constants of the thin film material to normalize the curve. We have elucidated the effects of size and profile of the x-ray beam and size of the sample on the geometrical factor. Using this method, geometrical factors could be successfully removed from the data to yield Fresnel reflectivity. Determining spill over angle was the key issue in determining the geometrical factor. The modus operandi described in this work for determining spill over angle by comparing usual experimental XRR profiles with the XRR profile taken with a surface contact knife edge is a heuristic solution that is well explained by intuitive modelling of the geometrical factor. This is the main aspect which makes it different and effective from the available literature. We have shown that theoretical Freshnel reflectivity R(θ) can never be extracted from the experimental reflectivity curve. Only R(θ) /R($θ_1$) can be extracted by using a legitimate data reduction procedure. Hence we propose to do away with the prevalent procedure of fitting the experimental reflectivity with R(θ) in the favour of fitting it with R(θ)/R($θ_1$), where $θ_1$ is the lowest angle after removing the direct beam part. The success of the method is demonstrated on $Er_2O_3$ thin film samples. Though it will be incorrect to draw any precise numerical connection (it is beyond the scope of this work) between the variation of lattice parameter found from XRD and the density of the material found from XRR, it is none the less the success of the method to detect such precise density variation when there is a change of lattice parameter.

**Acknowledgements**   Acknowledgements: The authors are thankful to Dr. P.A. Naik, director RRCAT for his support and encouragement during the course of this work.